\documentclass[12pt]{article}
\usepackage{amssymb}
\usepackage{amsmath}
\usepackage{graphicx,epstopdf}
\newcommand{\doublespacing}{\let\CS=\@currsize\renewcommand{\baselinesstrech}
{2.0}\tiny\CS}

\newcommand{\bd}{\begin{document}}
\newcommand{\ed}{\end{document}}
\newcommand{\bc}{\begin{center}}
\newcommand{\ec}{\end{center}}
\newcommand{\bfg}{\begin{figure}}
\newcommand{\efg}{\end{figure}}
\newcommand{\vs}{\vspace}
\newcommand{\beqas}{\begin{eqnarray*}}
\newcommand{\eeqas}{\end{eqnarray*}}

\begin{document}

\title {\Large \textbf {Dust-Ion-Acoustic Waves in unmagnetized 4-component plasma}}
\author{Anjana Sinha $^{1,a}$ and Biswajit Sahu$^{2,b}$ \\
1. Physics and Applied Mathematics Unit,\\ Indian Statistical Institute, Kolkata - 700 1 08, India \\
2. Department of Mathematics, \\West Bengal State University, Barasat, Kolkata-700126, India}


\date{}

\maketitle 


  \begin{center}
  	{\bf ABSTRACT}
  \end{center}

A theoretical study is presented for the propagation of Dust Ion Acoustic Waves in an unmagnetized four-component plasma, consisting of Maxwellian negative ions, cold mobile positive ions, $\kappa$-distributed electrons and positively charged dust grains. Based on the characteristics of Sagdeev pseudopotential and phase portraits,  three types of nonlinear waves are observed --- solitons, double layers and supersolitons. The conditions for the existence of such nonlinear waves are highly sensitive to the plasma parameters. The results obtained in this study may be of wide relevance in the field of space plasma as well as ultrasmall semiconductor devices in the laboratory.

  \thispagestyle{empty} 

  \setlength{\baselineskip}{18.5 pt}

\vspace{0.4 cm}


\vspace {.5 cm}

\noindent
$\overline{ {^a E-mail : sinha.anjana@gmail.com}}$ \\
${^b E-mail : biswajit_{-}sahu@yahoo.co.in}$ 

 
\pagebreak
\section{Introduction :}

The presence of charged dust grains in a plasma not only modifies the existing plasma wave spectra but also introduces a number of novel eigenmodes, such as the Dust-Ion-Acoustic (DIAWs) waves \cite{shukla-silin} --- a low frequency analogue of Ion Acoustic Waves --- observed both in laboratory \cite{nakamura1, nakamura2, barkan} and space environments \cite{space1, space2, space3, basudev, epjp-depsy}, such as cometary tails, planetary rings, supernovae, pulsars, cluster explosions, active galactic nuclei and interstellar medium.. Thus DIAWs play a vital role in understanding different types of collective processes in such dusty plasmas. On the other hand, super nonlinear solitons (or supersolitons) form an interesting class of special class of solitary waves, theoretically discovered fairly recently in the last decade \cite{dubinov1, dubinov2}. Theoretical studies involving multi-species plasmas containing three or more kinds of particles --- ranging from the lightest (neutrinos, electrons, positrons) to the heaviest ones (fullerene ions and charged dust grains) --- have yielded what are called supersolitons, attracting the attention of astrophysicists studying compact astrophysical objects like white dwarfs, neutron stars and black holes \cite{dubinov1}.   Mathematically, supersolitons are characterized by intriguing features --- the associated Sagdeev pseudopotential $V(\phi)$ has at least two local minima, separated by at least one local maximum, while remaining negative in this region. The minima correspond to the equilibrium points of the system, while the maximum, where the separatrix intersects itself, denotes the unstable equibrium or the saddle point. This leads to a characteristic supersoliton signature in the electric field --- the usual simple bipolar shape has an extra wiggle or fold on each side of the structure \cite{kourakis-pop2013}. As regards their phase portraits, these highly nonlinear periodic waves have trajectories enveloping the separatrix, thus indicating that their  total energy is above a certain potential
barrier height and the amplitude cannot be smaller than that of the separatrix. 
Analogous to solitons, these nonlinear structures propagate at fixed velocity, without changing form, but with amplitude and width much greater than those of regular solitons; hence they are termed as supersolitons.

 Interestingly, many such features had been
observed earlier \cite{ss-early1, ss-early2, ss-early3, ss-early4}, without being recognized as supersolitons. Typically, these supersolitary structures occur at wave speeds above those of double layers. Incidentally, till a few years ago, double layers were regarded as upper limits for the existence of solitons \cite{hellberg-2013}. However, supersolitons may appear even without a double layer \cite{steffy}. In a fairly recent work involving fluid simulation, it was shown that supersolitons are stable structures, thus increasing the possibility of their direct observation \cite{stable}. In passing, we note that Electron Acoustic supersolitary waves raise the possibility of their application to the interpretation of satellite observations of electric field data \cite{kamalam-steffy}.

Various scientists have studied nonlinear wave propagation in multi species plasma, considering different kinds of particles, with special emphasis on supersolitons. For example, super IAWs in non-magnetized  multispecies plasma containing electrons, positrons, and two types of ions of the
different signs of charge and dust particles \cite{dubinov-ieee2012}, electrons, protons and two types of ions \cite{dubinov1}, positive and negative charged dust species with non-extensively  distributed electrons and ions \cite{el-wakil2017}, negatively charged static dust grains, adiabatic warm ions, nonthermal electrons and isothermal / nonthermal positrons \cite{anup1,anup2}, to cite a few. Ion acoustic supersolitons are also observed in magnetized plasma, with a cold ion
fluid, cool Boltzmann electrons and nonthermal energetic hot electrons \cite{lakhina-pop2015,kamalam}. Positron-acoustic Gardner solitons and double layers in four-component plasma system consisting of immobile positive ions, mobile cold positrons, nonthermal hot positrons and nonthermal hot electrons were studied in ref. \cite{epjp-rahman}. Initially, it was conjectured that plasma should consist of at least four different types of particles for the existence of supersolitons \cite{dubinov1, dubinov2, dubinov-ieee2012}. However, later it was shown in a plasma consisting of cold positive and negative ions, with Cairns-distributed electrons, that the minimum number of species to support supersolitons is three \cite{verheest-2013a}. Subsequently studies on various plasma models, each with unique 3-species combintions, give credence to this fact \cite{verheest-2013b, verheest-2013c, maharaj-2013, verheest-2015, olivier-2017}. A review article by Dubinov and Kolotkov \cite{dubinov2018} is of particular mention in this regard. 

A sizeable number of studies consider Maxwell-Boltzmann distribution for the electrons and ions. However, in general, space plasmas have non Maxwellian distribution \cite{hellberg-2013}. Various interplanetary missions have confirmed the presence of nonthermal particle distribution in the solar wind and near Earth space plasma \cite{pierrard-2010, kappa-astro, kappa-book}. Cairns et. al. \cite{cairns} introduced a theoretical model to represent the effects of non thermal component, that has been used copiously in theoretical studies of nonlinear structures. In low density plasma in the Universe, where binary collisions of charges are sufficiently rare, the super thermal populations may well be described by the $\kappa$-velocity distribution function. This function has a high energy tail deviating from the Maxwellian distribution and decreasing as a power law in particle speed
\begin{equation}\label{kappa}
\displaystyle n_e = \displaystyle \left( 1 - \frac{\Phi}{\kappa - \frac{3}{2} } \right)^{ - \kappa + 1/2} 
\end{equation}
where $\Phi$ is the electrostatic wave potential normalized by $k_B T_e/e$, $k_B$ denoting the Boltzmann constant and $T_e$ representing the electron temperature. The $\kappa$ value indicates the spectral index, determining the slope of the energy spectrum of the particles forming the tail of the velocity distribution function. The critical value of $\kappa$ where the distribution function collapses and equivalent temperature is not defined, is $\kappa _c = 3/2$. So $\kappa > 3/2$. For $\kappa \rightarrow \infty$, the distribution reduces to the Maxwellian one. $2 < \kappa < 6$ fits observation from satellite data in solar wind, terrestrial magnetosphere, terrestrial plasma sheet, plasma sheets of Mercury, Jupiter, Saturn etc. Various studies have invoked different mechanisms to explain the physical origin of $\kappa$-distribution. Whatever the mechanism, this distribution is a powerful mathematical tool to fit observational data in space plasmas. Over the years $\kappa$ distribution has become increasingly widespread across astrophysical plsma processes, describing velocity and energy o particles from solar wind and planetary magnetospheres to heliosheath, beyond to interstellar and intergalactic plasmas \cite{kappa-book}. Recent advances in space physics show thw the connection of $\kappa$ distribution with statistical mechanics and thermodynamics, thus enabling us to improve our understanding on the statistical origin of $\kappa$ distrobution and develop the possible physical mechanisms that create the required statistical environment to obtain such distributions in the laboratory.

In this study we shall investigate the propagation of dust ion acoustic waves (DIAWs) in a collisionless unmagnetized four-species electron-ion dusty plasma, consisting of cold mobile inertial positive ions, Maxwellian negative ions, $\kappa$-distributed electrons and positively charged stationary dust grains. This model has applications not only in astrophysical environments, but also in ultra small semiconductor devices \cite{shukla-book}. In a recent study \cite{olivier-pop25}, the authors found the criteria for the existence of supersolitons, in a plasma consisting of cold ions and two-temperature Boltzmann electrons. Our main aim here is to look for small amplitude supersolitons in the present model. In particular, starting from the fluid equations, we shall apply the reductive perturbation technique to arrive at the Sagdeev pseudopotential. We shall also plot the pseudopotential, along with corresponding phase portraits, and search for nonlinear solitary structures like solitons, double layers and supersolitons. To ascertain their structure, we shall also plot the associated electric field profile.

The paper is structured as follows : The basic governing equations are given in Section 2. To have an idea of the linearized waves, the linear dispersion relation is derived in Section 3. The reductive perturbative analysis is carried out in Section 4, to arrive at an expressioin for the pseudopotential. Whether this model supports supersolitons following the criteria given in ref. \cite{olivier-pop25} is investigated in Section 5. Section 6 shows the various plots for the pseudopotential, phase portraits, wave structure and eletric fields, to have an idea of the propagating Dust Ion Acoustic waves. Finally, Section 7 is kept for Conclusions and Remarks.

\section{Nonlinear Governing Equations}

As mentioned earlier, in this study we shall investigate the propagation of dust ion acoustic waves (DIAWs) in a collisionless four-species electron-ion dusty plasma, consisting of cold mobile inertial positive ions, Maxwellian negative ions, $\kappa$-distributed electrons and positively charged stationary dust grains. Such a mathematical model of four-component dusty
plasma has a significant importance in different space
media such as upper mesosphere, comets tails, etc., where the dust in space may be
charged due to the effects of thermionic secondary or photo
emissions \cite{el-wakil2017, rahman}. The basic normalized fluid equations for the dust ion acoustic waves are given by
\begin{equation}\label{gov1}
\displaystyle \frac{\partial n_p}{\partial t} + \frac{\partial}{\partial x} \left( n_p u_p \right) = 0
\end{equation}
\begin{equation}\label{gov2}
\displaystyle \frac{\partial u_p}{\partial t} + u_p \frac{\partial u_p}{\partial x}  = - \frac{\partial \Phi }{\partial x}
\end{equation}
along with the Poisson equation
\begin{equation}\label{gov3}
\displaystyle \frac{\partial ^2 \Phi}{\partial x^2}   = \displaystyle \mu _e \left( 1 - \displaystyle \frac{\Phi}{\kappa - \frac{3}{2}} \right)^{\frac{1}{2} - \kappa} + \mu _n e^{\alpha \Phi} -n_p - \mu _d
\end{equation}
where the different notations denote the following : $u_p$ is the positive ion fluid speed, 
$\Phi$ is the normalized electrostatic potential, $n_{e0}$, $n_{p0}$ and $n_{d0}$ represent the equilibrium value number densities of electron, positive ion and dust, respectively, $\mu _e = n_{e0}/n_{p0}$, $\mu _d = \displaystyle \frac{Z_d n_{d0}}{n_{p0}} $, with $Z_d$ being the charge on the dust grains, $\alpha = T_e / T_n$, $T_e$ and $T_n$ representing the electron and negative ion temperatures, respectively. It is worth mentioning here that space is normalized to the Debye wavelength $\lambda _D = \displaystyle \sqrt{ \frac{\varepsilon _0 k_b T_e}{n_{p0} e^2}}$, and time is normalized to the plasma period for positive ions $\omega _{pp} ^{-1} = \displaystyle \left( \frac{n_{p0} e^2 }{\varepsilon_0 m_{ip}} \right)^{-1/2} $, $m_{ip}$ being the mass of positive ion. Additionally, all number densities are normalized with respect to the equilibrium density for positive ion $n_{p0}$, the fluid velocity $u_p$ is normalized with respect to the ion-acoustic speed for positive ions
$c_{ip} = \sqrt{k_B T_{e} / m_{ip}} $, and the electrostatic potential is normalized by $\displaystyle \left( k_B T_e /e \right)$.
The charge neutrality condition at equilibrium demands 
\begin{equation}\label{charge-neutrality}
n_{p0}+Z_d n_{d0}= n_{e0} +n_{n0}
\end{equation}
so that $\mu _n = 1 - \mu _e + \mu _d$. 
Our next step would be to solve the equations (\ref{gov1}) to (\ref{gov3}), to have an insight into the propagation of Dust Ion Acoustic waves in the medium. But before that we would just like to have an idea about the dispersion relation of the corresponding linear waves.

\vspace{1cm}

\section{Linear Analysis}

In order to deduce the linear dispersion relation for the plasma waves, we assume the following expansions
\begin{equation}\label{expand}
n _{p,e} = 1 + n_{p,e} ^{(1)} \qquad , \qquad \Phi = \Phi ^{(1)} \qquad , \qquad u_{p} =  u_{p} ^{(1)}
\end{equation}
with the perturbations $n_{p,e} ^{(1)}, \ \Phi ^{(1)} $ and $u_{p} ^{(1)}$ proportional to $\displaystyle e^ {i( k x - \omega t)}$. Here $\omega$ and $k$ are the wave frequency and wave vector, respectively. After some straightforward algebra, we get the dispersion relation as :
\begin{equation}\label{dispersion}
\omega = \displaystyle \frac{k}{\sqrt{k^2 + \mu _n \alpha + \mu _e \left(\frac{\kappa - 1/2}{\kappa - 3/2}\right)}}
\end{equation}
Thus the linear dispersion relation of the Dust Ion Acoustic waves is quite a simple one. With the help of Mathematica, we plot the behaviour of $\omega$ wrt $k$ in Fig. 1. For very small wave vector, the frequency, $\omega$, of the linear waves varies linearly with $k$, and finally reaches unity. The spectral index $\kappa$ has very little influence on the linear dispersion relation --- for very low values of $\kappa$, the wave frequency $\omega$ reaches unity at a slightly greater value of $k$.

\bfg
\bc
\includegraphics[width = 11 cm]{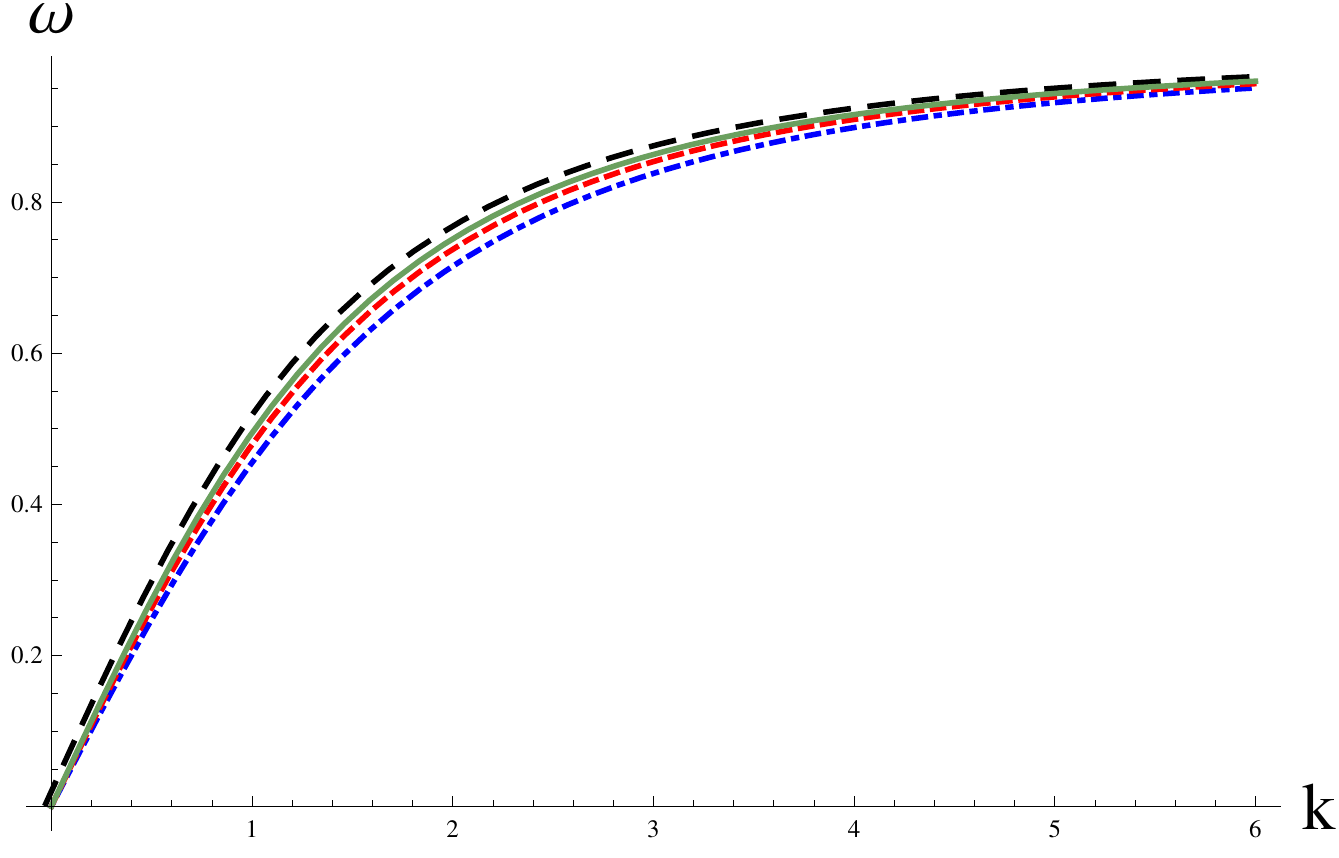}  \\
\caption{Color Online :Plot of $\omega$ vs $k$, for different $\kappa$. The uppermost black dashed curve is for $\kappa = 4.1$, the lowermost dotdashed blue curve is for $\kappa = 2.1$. The solid green curve (second from above) and the dotted red curve (third from above) are for $\kappa =3.12 $ and $2.5 $, respectively. Other plasma parameters are $\mu _e = 0.7, \ \mu _d = 0.19, \  \alpha = 4$ }
\ec
\efg

\section{Reductive Perturbative Analysis}

Since our main aim in this study is to look for small amplitude supersolitons, we shall consider plasma near ‘supercritical’ composition \cite{verheest-jpp2015, verheest-jpp2016}.  The main reason behind this is the study by Olivier et. al., which showed that supercritical plasma compositions are related to small amplitude supersolitons \cite{olivier-2017}. At ‘supercritical’ plasma composition, the nonlinear terms in both the resulting KdV and mKdV
equations vanish simultaneously for the same set of plasma parameters. For this purpose, we need to retain fourth order nonlinear effects in the reductive perturbation expansion \cite{olivier-pop25}. So we introduce stretched coordinates of the form :
\begin{equation}\label{rpt-coord}
\xi = \epsilon ^{3/2} (x-M t) \qquad , \qquad \tau = \epsilon ^{9/2} t
\end{equation}
where $M$ is the linear acoustic phase velocity, and $\epsilon$ determines the strength of nonlinearity. All physical quantities can be expanded in terms of power series in $\epsilon$, about their equilibrium values, as
\begin{equation}\label{rpt}
\begin{array}{lll}
n_{p} &=& \displaystyle 1 + \epsilon n_{p} ^{(1)} + \epsilon ^2 n_{p} ^{(2)} + \epsilon ^3 n_{p} ^{(3)} + \epsilon ^4 n_{p} ^{(4)} + \cdots \\ \\
\Phi &=& \displaystyle  \epsilon \Phi ^{(1)} + \epsilon ^2 \Phi ^{(2)} + \epsilon ^3 \Phi ^{(3)} + \epsilon ^4 \Phi ^{(4)} + \cdots \\ \\
u_{p} &=& \displaystyle  \epsilon u_p ^{(1)} + \epsilon ^{2} u_{p} ^{(2)} + \epsilon ^3 u_{p}  ^{(3)} + \epsilon ^4 u_{p} ^{(4)} + \cdots \\ \\
\end{array}
\end{equation}
with boundary conditions $n_p \rightarrow 1, \ u_p \rightarrow 0, \ \Phi \rightarrow 0$ as $| \xi | \rightarrow \infty$. 
Plugging in eqns. (\ref{rpt-coord}) and (\ref{rpt}) in eqns. (\ref{gov1}) to (\ref{gov3}), we obtain the different equations in various orders of $\epsilon$. In particular, the continuity equation yields
\begin{equation}\label{order-cont}
\begin{array}{lll}
\displaystyle n_{p} ^{(1)} &=& \displaystyle \frac{1}{M } u_p ^{(1)}  \\ \\
\displaystyle n_p ^{(2)} &=& \displaystyle \frac{1}{M} \left\{ u_p ^{(2)} + n_p ^{(1)} u_p ^{(2)} \right\} \\ \\
\displaystyle n_{p} ^{(3)} &=& \displaystyle \frac{ 1 }{M} \left\{ u_p ^{(3)} + n_p ^{(1)} u_p  ^{(2)} + n_p ^{(2)} u_p ^{(1)} \right\} \\ \\
\displaystyle n_p ^{(4)}  &=& \displaystyle  \frac{1}{M} \left\{ u_p ^{(4)} + n_p ^{(1)} u_p ^{(3)} + n_p ^{(2)} u_p ^{(2)} + n_p ^{(3)} u_p ^{(1)} \right\}   \\ \\
\end{array}
\end{equation}
whereas the momentum equation gives
\begin{equation}\label{order-mom}
\begin{array}{lll}
\displaystyle u_{p} ^{(1)} &=& \displaystyle \frac{1}{M } \Phi ^{(1)}  \\ \\
\displaystyle u_p ^{(2)} &=& \displaystyle \frac{1}{M} \left\{ \Phi ^{(2)} + \frac{1}{2} \left(  u_p ^{(1)} \right) ^2 \right\} \\ \\
\displaystyle u_{p} ^{(3)} &=& \displaystyle \frac{ 1 }{M} \left\{ \Phi ^{(3)} + u_p ^{(1)} u_p  ^{(2)}  \right\} \\ \\
\displaystyle \frac{\partial u_p ^{(4)}}{\partial \xi}  &=& \displaystyle  \frac{1}{M} \left\{ \frac{\partial }{\partial \xi} \Phi ^{(4)} + u_p ^{(1)} \frac{\partial u_p ^{(3)}}{\partial \xi} + \frac{\partial u_p ^{(1)}}{\partial \xi} u_p ^{(3)} + u_p ^{(2)} \frac{\partial u_p ^{(2)} }{\partial \xi} \right\} + \frac{1}{M} \frac{\partial u_p ^{(1)}}{\partial \tau}  \\ \\
\end{array}
\end{equation}
Combining equations (\ref{order-cont}) and (\ref{order-mom}), gives the number density of the positive ion in terns of the leading order pseudo potential. 
\begin{equation}\label{order-np}
\begin{array}{lll}
\displaystyle n_{p} ^{(1)} &=& \displaystyle \frac{1}{M } u_p ^{(1)} = \frac{1}{M^2} \Phi ^{(1)}  \\ \\
\displaystyle n_p ^{(2)} &=& \displaystyle \frac{1}{M^2} \Phi ^{(2)} + \frac{3}{2 M^4} \left(  \Phi ^{(1)} \right)^2 \\ \\
\displaystyle n_{p} ^{(3)} &=& \displaystyle \frac{ 1 }{M^2 } \Phi ^{(3)} + \frac{3}{M^4} \Phi ^{(1)} \Phi ^{(2)} + \frac{5}{2 M^6} \left( \Phi ^{(1)} \right) ^3 \\ \\
\displaystyle \frac{\partial n_p ^{(4)}}{\partial \xi}  &=& \displaystyle  \frac{\partial}{\partial \xi} \left\{ \frac{1}{M^2} \Phi ^{(4)} + \frac{35}{8 M^8} \left( \Phi ^{(1)} \right) ^4 +\frac{3}{M^4} \Phi ^{(1)} \Phi ^{(3)} + \frac{3}{2 M^4} \left( \Phi ^{(2)} \right) ^2 \right. \\ \\ 
& & \displaystyle + \left. \frac{15}{2 M^6} \left( \Phi ^{(1)} \right) ^2 \Phi ^{(2)}   \right\} + \frac{2}{M^3} \frac{\partial \Phi ^{(1)}}{\partial \tau} \\ \\
\end{array}
\end{equation}
Next we perform Taylor series expansion of the Poisson equation (\ref{gov3}) in terms of the stretched coordinates in eq. (\ref{rpt-coord}), retaining terms up to order $\epsilon ^4$ :
\begin{equation}\label{poisson-taylor}
\begin{array}{lll}
	& & \displaystyle \epsilon ^4 \Phi _{\xi \xi} ^{(1)} + \left[1 + \mu _d - \mu _e - \mu _n \right] + \epsilon \left( n_p ^{(1)} - A_1 \Phi ^{(1)} \right) + \epsilon ^2 \left[n_p ^{(2)} - A_1 \Phi ^{(2)} - \frac{A_2}{2} \left( \Phi ^{(1)} \right) ^2 \right] \\ \\ 
& & \displaystyle  + \epsilon ^3 \displaystyle \left[ n_p ^{(3)} - A_1 \Phi ^{(3)} - A_2 \Phi ^{(1)} \Phi ^{(2)} - \frac{A_3}{6} ( \Phi ^{(1)} )^3 \right] \\ \\ 
& & \displaystyle + \epsilon ^4 \left[ n_p ^{(4)} - A_1 \Phi ^{(4)} - \frac{A_2}{2} \left( \Phi ^{(2)} \right) ^2 -  A_2 \Phi ^{(1)} \Phi ^{(3)}  - \frac{A_4}{24} \left( \Phi ^{(1)} \right) ^4 -  \frac{A_3}{2} \left( \Phi ^{(1)} \right) ^2 \Phi ^{(2)} \right] = 0 \\ \\ 
\end{array}
\end{equation}
where each subscript $\xi$ denotes differentiation wrt $\xi$. In eq. (\ref{poisson-taylor}) above, 
\begin{equation}\label{a_j}
A_j = \displaystyle \alpha ^j \mu _n + \mu _e \frac{1}{j ! \ \displaystyle \left( \kappa - \frac{3}{2} \right) ^j}  \displaystyle \prod _{i=0} ^{j-1} \left( \kappa - i - \frac{1}{2} \right) 
\end{equation}
Putting eq. (\ref{order-np}) in eq. (\ref{poisson-taylor}), and equating each power of $\epsilon$ to zero, helps us to determine the different coefficients $A_j$ for the supercritical plasma composition. For instance, the zeroth order in $\epsilon$ yields the charge neutrality condition at equilibrium, viz., $ n_{p0} + Z_d n_{d0} =n_{e0} + n_{n0} $. The first, second and third orders in $\epsilon$ give 
\begin{equation}\label{a1-super}
 \displaystyle \left( A_1 - \frac{1}{M^2} \right) \Phi ^{(1)} = 0 
 \end{equation}
 \begin{equation}\label{a2-super} 
 \displaystyle \left( A_1 - \frac{1}{M^2} \right) \Phi ^{(2)} + \frac{1}{2} \left(  A_2 - \frac{3}{ M^4} \right) \left( \Phi ^{(1)} \right) ^2 = 0 
 \end{equation}
 \begin{equation}\label{a3-super} 
 \displaystyle \left( A_1 - \frac{1}{M^2} \right) \Phi  ^{(3)} + \left(  A_2 - \frac{3}{M^4} \right) \Phi ^{(1)} \Phi ^{(2)} + \frac{1}{2} \left( \frac{A_3}{3} -\frac{5}{M^6} \right) ( \Phi ^{(1)} )^3 = 0   
\end{equation}
Now $\Phi ^{(1)} \neq 0$, hence the coefficient of $\Phi ^{(1)}$ must vanish, so that $A_1 = 1/M^2$. This gives the dispersion law \cite{verheest-jpp2016}. Putting the coefficients of $ \displaystyle \left( \Phi ^{(1)} \right)^2$ and $ \displaystyle \left( \Phi ^{(1)} \right)^3  $ to zero amounts to annulment of the quadratic and cubic nonlineraties in the final equation for solitary waves, so that we are left with quartic nonlinearity.  Thus $A_1 = 1/M^2, \ A_2 = 3/M^4, \ A_3 = 15/M^6 $, at the supercritical plasma composition. However, we shall look for nonlinear structures near the supercritical composition. So we take the $A_1, \ A_2, \ A_3 $ values as follows :
\begin{equation}\label{a123}
\begin{array}{lcl}
\displaystyle A_1 &=& \displaystyle \frac{1}{ M^2} \\ \\
\displaystyle A_2 &=& \displaystyle \frac{3}{M^4} - \epsilon ^2 B_2 \ {\rm{(say) }} \\ \\
\displaystyle A_3 &=& \displaystyle \frac{15}{M^6} - \epsilon B_3 \ {\rm{(say) }} \\ \\
\end{array}
\end{equation}
with $B_2$, $B_3$ being independent of $\epsilon$. Differentiating eq. (\ref{poisson-taylor}), and using eq. (\ref{order-np}), the final eq. is obtained as
\begin{equation}\label{final-0}
\begin{array}{lll}
& & \displaystyle \epsilon ^4 \Phi _{\xi \xi \xi} ^{(1)} + \displaystyle \frac{2}{M^3} \epsilon ^4 \Phi _{\tau} ^{(1)} + \left[ \epsilon \left\{ \frac{1}{M^2} - A_1 \right\} \Phi ^{(1)} + \epsilon ^2 \left\{ \frac{1}{M^2} - A_1 \right\} \Phi ^{(2)}  \right. \\ \\    
& & \displaystyle  \left.  + \epsilon ^2 \left\{ \frac{3/M^4 - A_2}{2} \right\} \left( \Phi ^{(1)} \right) ^2 + \epsilon ^3 \left\{ \frac{1}{M^2} - A_1 \right\} \Phi ^{(3)}  + \epsilon ^3 \left\{ \frac{3}{ M^4} - A_2 \right\}  \Phi ^{(1)} \Phi ^{(2)}   \right. \\ \\
& & \displaystyle \left. + \epsilon ^3 \left\{ \frac{15 /  M^6 - A_3 }{6} \right\} \left( \Phi ^{(1)} \right) ^3  + \epsilon ^4 \left\{ \frac{1}{M^2} - A_1 \right\} \Phi ^{(4)} 
 + \epsilon ^4 \left\{ \frac{3 / M^4 - A_2}{2} \right\} \left(\Phi ^{(2)} \right) ^2 
 \right. \\ \\ 
& & \displaystyle \left.   + \epsilon ^4 \left\{ \frac{3}{M^4} - A_2 \right\} \Phi ^{(1)} \Phi ^{(3)} + \epsilon ^4 \left\{ \frac{105/M^8 - A_4}{24} \right\} \left( 
\Phi ^{(1)} \right) ^4  \right. \\ \\ 
& & \displaystyle \left. + \epsilon ^4 \left\{ \frac{15 / M^6 - A_2}{2}  \right\} \left( \Phi ^{(1 )} \right) ^2 \Phi ^{(2)} \right] _{\xi} = 0 \\ 
\end{array}
\end{equation}
Applying conditions in eq. (\ref{a123}) and retaining terms up to order $\epsilon ^4$, eq. (\ref{final-0}) reduces to an expression for the lowest order of the electrostatic potential $\Phi ^{(1)}$ : 
\begin{equation}\label{final}
\displaystyle \Phi _{\tau} ^{(1)} + M^3 \left[ \frac{1}{2}  \Phi _{\xi \xi \xi} ^{(1)} + \displaystyle \frac{B_2}{2} \Phi ^{(1)} \Phi _{\xi} ^{(1)}   + \frac{B_3}{4} \left( \Phi ^{(1)} \right) ^2 \Phi ^{(1)} _{\xi} + \frac{( 105/M^8 - A_4 ) }{12} \left( \Phi ^{(1)} \right) ^3 \Phi ^{(1)} _{\xi}  \right] = 0
\end{equation}
Going back to the original coordinates
\begin{equation}
t = \displaystyle \epsilon ^{-9/2} \tau \qquad , \qquad \eta = \displaystyle \left( x - M t \right) = \epsilon ^{-3/2} \xi
\end{equation}
and expressing the electrostatic potential $\Phi$ in terms of its lowest order, viz., $\Phi = \epsilon \Phi ^{(1)}$, eq. (\ref{final}) can be cast in the form 
\begin{equation}\label{phi-eta}
\displaystyle \Phi _{\tau}  + M^3 \left[ \frac{1}{2}  \Phi _{\eta \eta \eta}  + \displaystyle a \Phi \Phi _{\eta}   + b  \Phi  ^2 \Phi _{\eta} + c \Phi ^3 \Phi  _{\eta}  \right] = 0
\end{equation}
where the coefficients $a, \ b, \ $ stand for 
\begin{equation}\label{abc}
a = \displaystyle \epsilon ^2 \frac{B_2}{2} \qquad b = \displaystyle \epsilon \frac{B_3}{4} \qquad c = \displaystyle \frac{105 / M^8 - A_4}{12} 
\end{equation}
We now try to find wave solutions to the final eq. (\ref{phi-eta}) in the next section.

\section{Criteria for existence of supersolitons}

To find wave solutions to eq. (\ref{phi-eta}), propagating with speed $v$, we introduce a moving reference frame
\begin{equation}\label{zeta}
\zeta = \eta - v t
\end{equation}
$v$ is also referred to as the Mach number. Henceforth we put $M=1$ for ease of calculations. 
Plugging in eq. (\ref{zeta}) in eq. (\ref{phi-eta}), and integrating the resulting equation twice, we arrive at an energy-like equation
\begin{equation}\label{dv-dphi}
\displaystyle \frac{1}{2}\frac{d}{d \zeta} \Phi ^2 + V (\Phi) = 0
\end{equation}
where the Sagdeev pseudopotential $V(\Phi)$ is expressed as
\begin{equation}\label{pseudopot}
V(\Phi) = \displaystyle - v \Phi ^2 + \frac{a}{3} \Phi ^3 + \frac{b}{6} \Phi ^4 + \frac{c}{10} \Phi ^4
\end{equation}
For supersolitons to exist, the Sagdeev pseudopotential must have three charge-neutral points $\Phi _c$, of the same sign, such that $\partial V / \partial \Phi = 0$ and $V<0$ at $\Phi = \Phi _c$. 
\begin{equation}\label{phi-s}
\displaystyle \frac{\partial V}{\partial \Phi} = \Phi \left\{ 2 v + a \Phi + \frac{2}{3}b \Phi ^2 + \frac{c}{2} \Phi ^3 \right\} = \Phi \ P_3 (\Phi) \ \ {\rm{(say)}}
\end{equation}
Now the roots of the polynomial $P_3 (\Phi)$ in eq. (\ref{phi-s}) will give the charge neutral points. Evidently, $\Phi = 0$ is the charge-neutral point at equilibrium. Hence for supersolitons to exist, the cubic polynomial $P_3 (\Phi)$ must have three distinct real roots with the same sign. In ref. \cite{olivier-2017}, the authors have worked out in detail the existence criteria for supersolitons. In particular they showed that supersolitons can exist only if the parameters $a, \ b, \ c$ satisfy the following relation :
\begin{equation}\label{abc-ss}
b < 0 \ , \qquad ac > 0 \ , \qquad ac < \displaystyle \frac{8}{27} b^2
\end{equation}
Furthermore, it is also proved that in such a case, supersolitons exist in a short range of velocities 
\begin{equation}\label{v-ss}
\displaystyle v _{min} < v < v_{max}
\end{equation}
where the lower and upper bounds of velocity are given by :
\begin{equation}\label{v-max}
\displaystyle v_{max} = v_{+}
\end{equation}
\begin{equation}\label{v-min}
\displaystyle v_{min} = \displaystyle \left\{ 
\begin{array}{lcl}
& & \displaystyle v_{DL} \ \ {\rm{if}} \ \  \frac{ac}{b^2} < \frac{5}{18}  \\ \\ 
& & \displaystyle v_{-} \ \ \ \ {\rm{if}} \ \ \frac{5}{18} < \frac{ac}{b^2} < \frac{8}{27} \\ \\
\end{array}
\right.
\end{equation}
where $v_{DL}$ corresponds to the velocity of a double layer 
\begin{equation}\label{v-dl}
v_{DL} = \displaystyle \frac{\displaystyle 5 b \left( \frac{5 b^2 - 27 ac}{27}  \right) - 200 \left( \frac{ 5 b^2 - 18  a c }{180}  \right) ^{3/2}  }{27 c^2}
\end{equation}
and $v_{\pm}$ is given as
\begin{equation}\label{v-pm}
v_{\pm} = \displaystyle \frac{\displaystyle \frac{2 b}{27} \left( 16 b^2 - 81 a c \right) \pm 
\left( \frac{8 b^2 - 27 a c }{18} \right) ^{3/2} }{27 c^2}
\end{equation}
In passing we note that we have included equations (\ref{abc-ss}) to (\ref{v-pm}) from ref. \cite{olivier-pop25} to make our present study self contained.

\section{Numerical Plots :}

Using Mathematica, we plot the Sagdeev pseudopotential $V(\Phi)$, the corresponding phase portraits, the electric field $E$ and the Dust Ion Acoustic wave for different values of plasma parameters, obeying criteria (\ref{abc-ss}) for the existence of supersolitons. It is worth mentioning here that the phase-portrait approach, based on the analysis of separatrices, is a powerful tool to identify different types of nonlinear waves. 

In particular, in Fig. 2 (a), we plot the pseudopotential $V(\Phi)$ vs $\Phi$ for different values of the dust concentration $\mu _d$, and in Fig. 2 (b), the phase portraits for the corresponding values of the plasma parameters. It is observed that within permissible limits, relatively lower values of dust concentration gives supersolitons, as shown by the bottom most dashed red curve ($ \mu _d = 0.188 $) and the next higher dotdashed green curve ($\mu _d = 0.189$). At some critical value of $\mu _d$, viz. $\mu _{dc} = 0.19014$ the solution is a double layer (shown by the solid blue curve, third from bottom), beyond which the solutions are solitons (depicted by the uppermost black curve, with $\mu _d = 0.192$). The corresponding trajectories in the phase space, as depicted in Fig. 2 (b), are self explanatory. The closed innermost orbit, represented by the solid black curve, signifies the regular soliton solution, whereas the separatrix curve shown in blue refers to the double layer. This trajectory has two centers corresponding to the two minima in the potential curve. On the other hand, the dotdashed green and the dashed red closed curves, around that of the double layer separatrix, represent supersolitons. The closed trajectories (of the supersolitons) in the phase portrait indicates the existence of an invariant energy in the dynamical system. This is often referred to as a generalised potential energy, which has a minimum coinciding with the position of the stable equilibrium point \cite{dubinov2018}. The amplitudes of these supersolitons can exceed the size of separatrix curves unlike
the regular soliton. This is clear in Fig. 3. The soliton solution corresponding to the plasma parameter values in Fig. 2 is represented by the black curve, even in Fig. 3, while the dashed red curve in Fig. 2 yields the supersoliton. The amplitudes of the soliton and the supersoliton in Fig. 3 match with those in Fig. 2 (b). In Fig. 4 (a) and 4 (b), we plot the electric field profiles for the corresponding supersoliton and soliton.  As expected, the electric field of the supersoliton has an extra wiggle (Fig. 4(a)).

\bfg
\bc
\includegraphics[width = 7 cm]{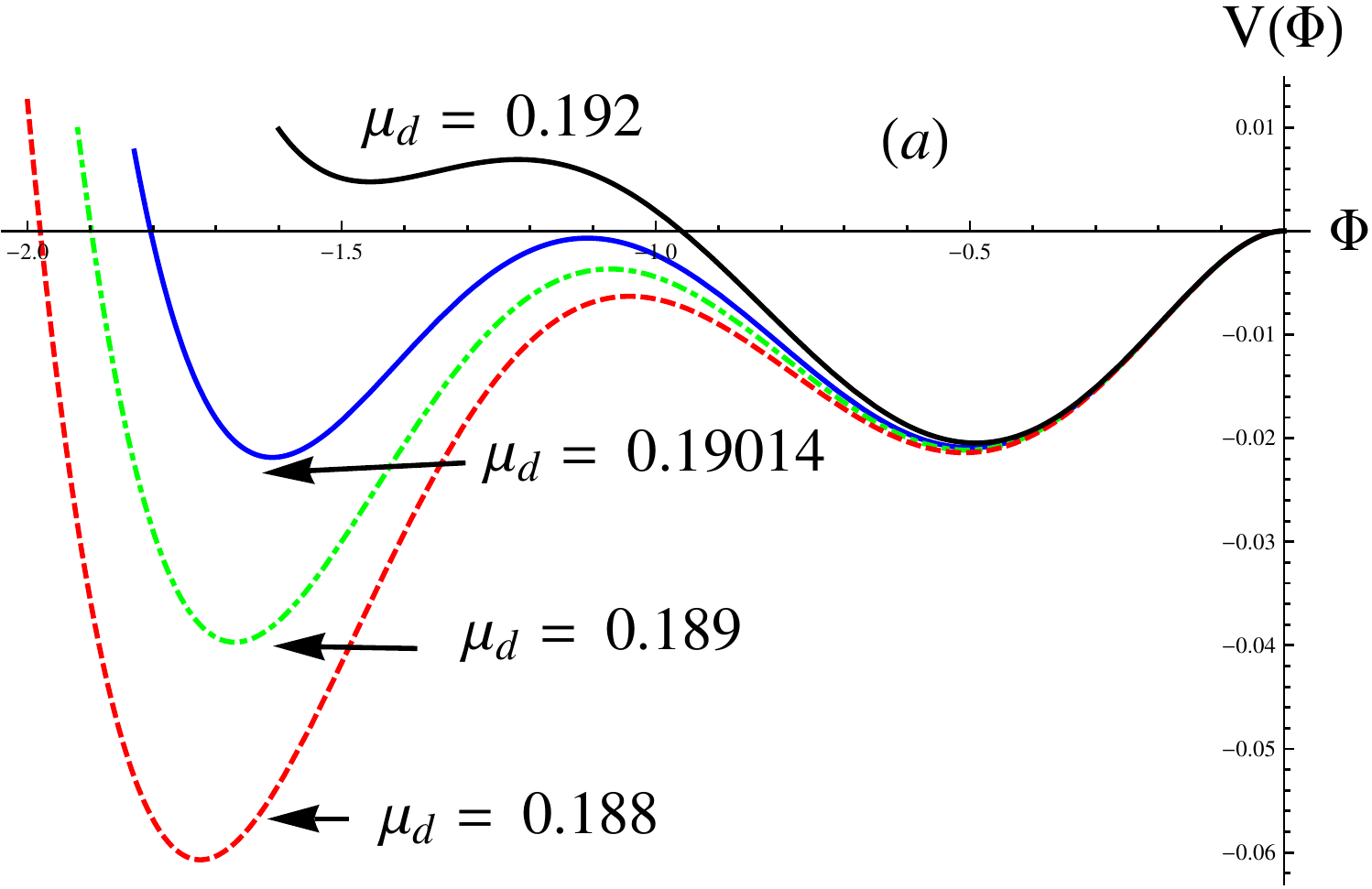} ~~ \includegraphics[width = 7 cm]{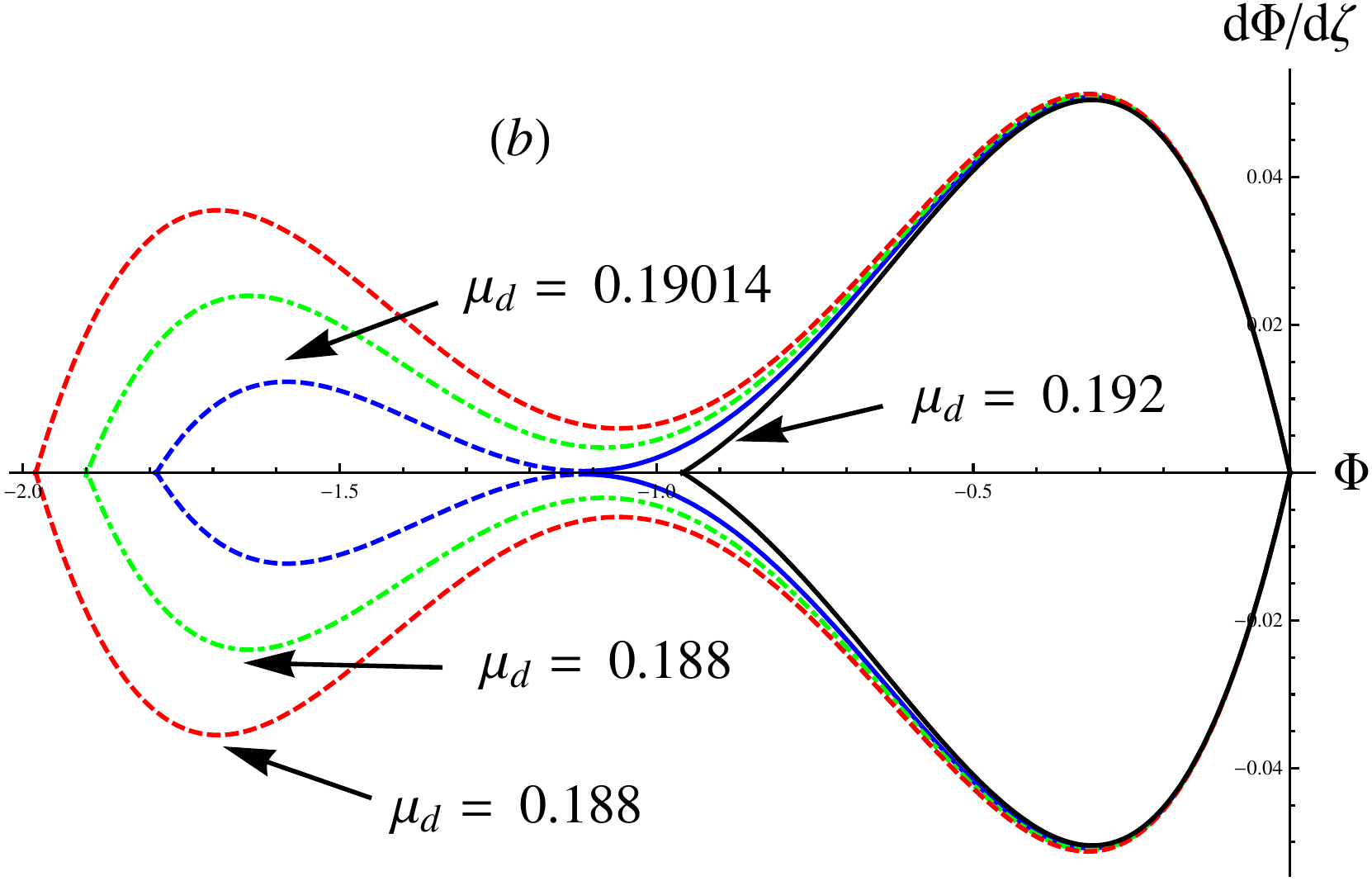} \\
\caption{(a) : Plot of $V(\Phi)$ vs $\Phi$, for different concentrations of dust grain, for $\mu _e = 0.7, \ \kappa = 3.6, \ \alpha = 4, \ v = 0.38$  ~~ (b) : Plot of $d \Phi / d \zeta$ vs $\Phi$ for corresponding different concentrations of dust grain. The black curves are for the highest value of dust concentration ($\mu _d = 0.192 $) and the red ones for the lowest value ($\mu _d = 0.188$). The black curves depict a soliton, the blue ones ($\mu _d = 0.19$) show a double layer and the green and red ones show supersolitons. }
\ec
\efg

\bfg
\bc
\includegraphics[width = 10 cm]{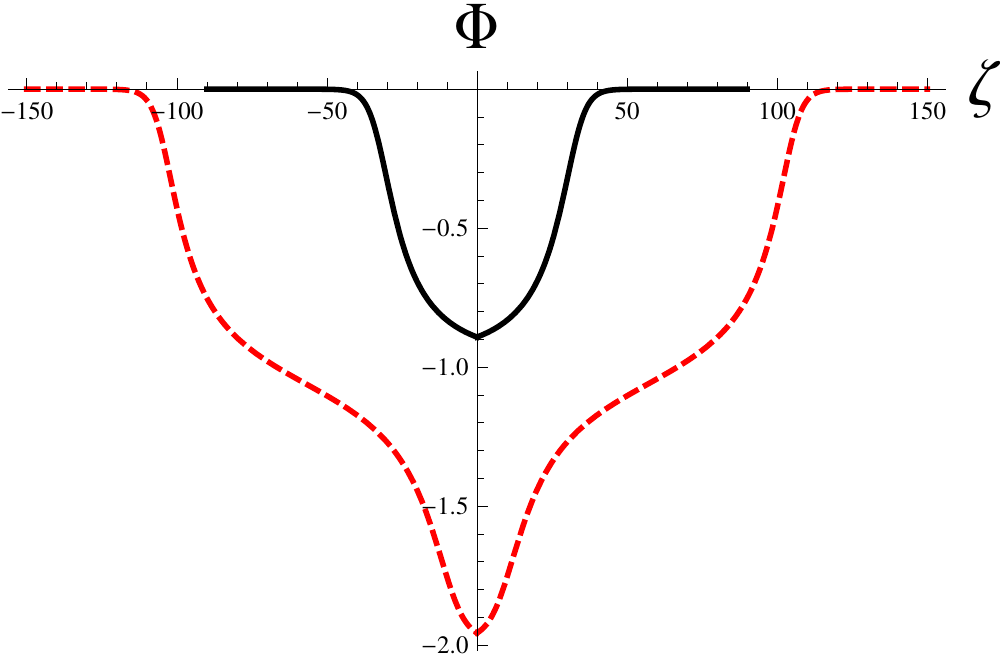} \\
\caption{Plot showing supersoliton in dotted red and soliton in solid black, for plasma parameter values for the red and black curves in Fig. 1, respectively. }
\ec
\efg

\bfg
\bc
\includegraphics[width = 7 cm]{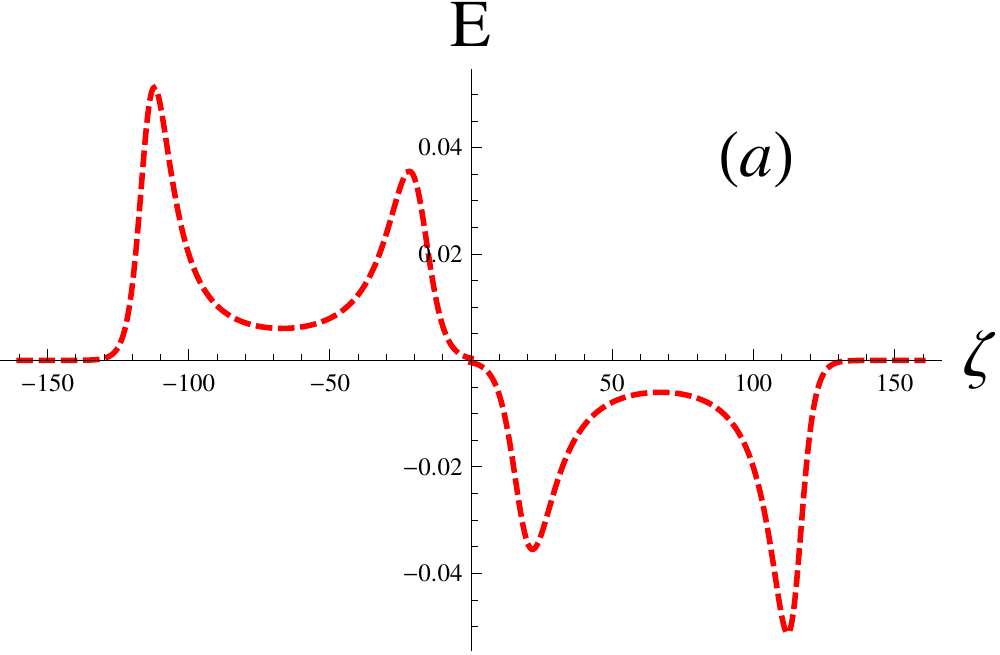} ~~ \includegraphics[width = 7 cm]{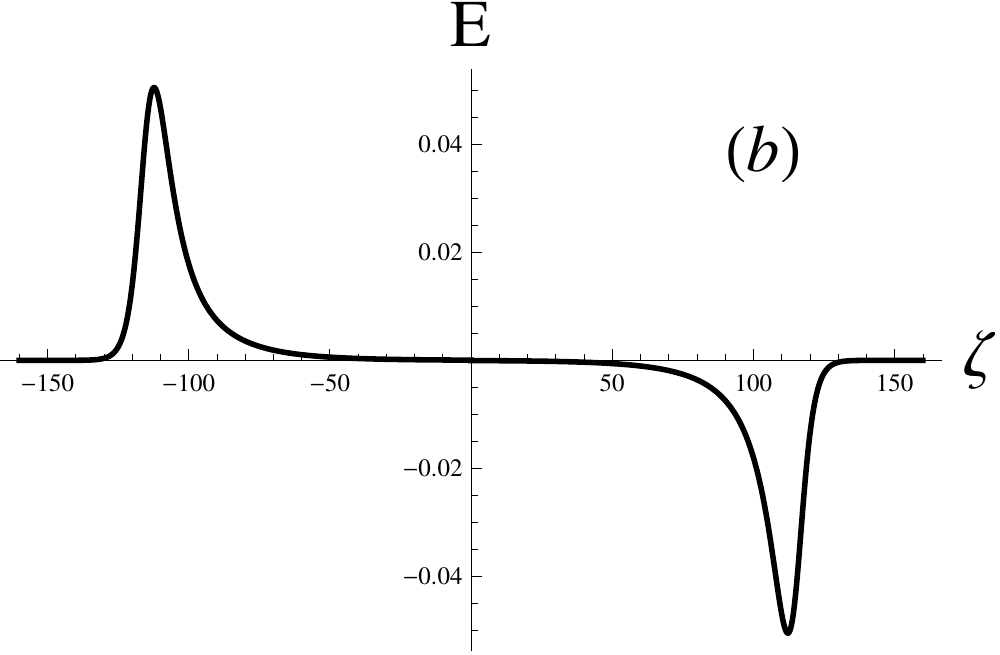} \\
\caption{(a) : Plot of the electric field $E$ vs $\zeta$, for supersoliton, for the plasma parameter values for the red dotted curves in Fig. 1. ~~ (b) : Plot of $E$ vs $\zeta$ for soliton, for the plasma parameter values for the solid black curve in Fig. 1 }
\ec
\efg

In Fig. 5 (a), we plot the pseudopotentials for varying $\kappa$ values, keeping other plasma parameters constant, and in Fig. 5 (b), the corresponding phase portraits. Comparatively smaller $\kappa$ value gives a soliton solution shown by the uppermost solid black plot in Fig. 5(a) and innermost closed black curve in Fig. 5 (b). A critical value $\kappa _c = 3.12$ yields a double layer (solid magenta curve --- second from above in Fig. 5a and the separatrix in Fig. 5b), while relativdely higher values of $\kappa$ support supersolitons (dot dashed green, dotted red and dashed blue curves). In Fig. 6 we plot the pseudopotential and phase portraits for different values of the Mach number $v$. Once again we observe that the system supports three types of nonlinear waves depending on the Mach number, provided it remains within permissible limits --- soliton for $v=0.37$ shown by the solid black curves, double layer for a critical value $v _c = 0.3796$ shown by the dotted blue curves, and supersolitons for $v = 0.39$ and $0.396$, shown by the dashed red and dotdashed green curves respectively. Plotting the corresponding solitary waves and the electric fields give results analogous to Figures 3 and 4 respectivly; so are not included here.

\bfg
\bc
\includegraphics[width = 7 cm]{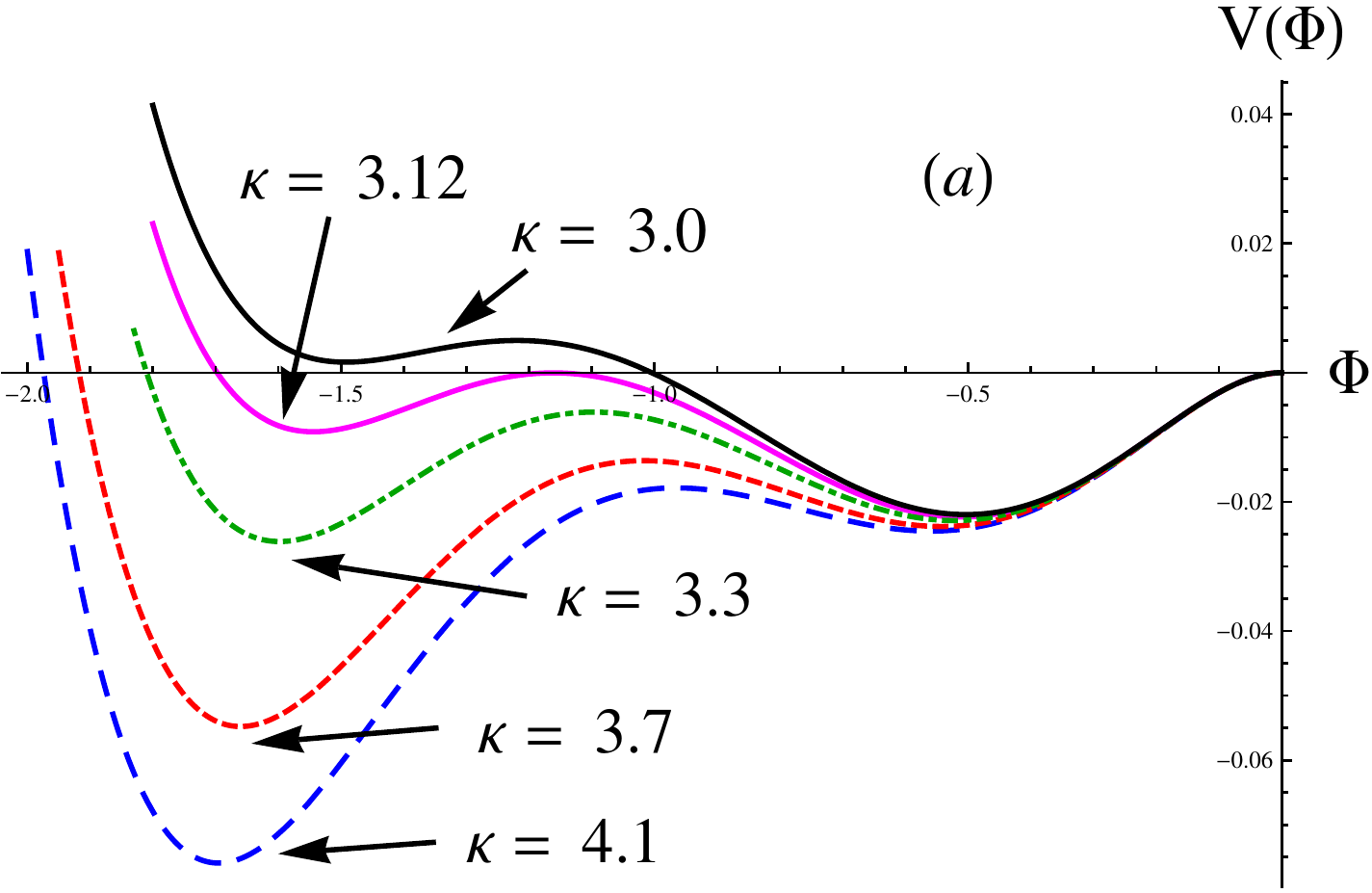} ~~ \includegraphics[width = 7 cm]{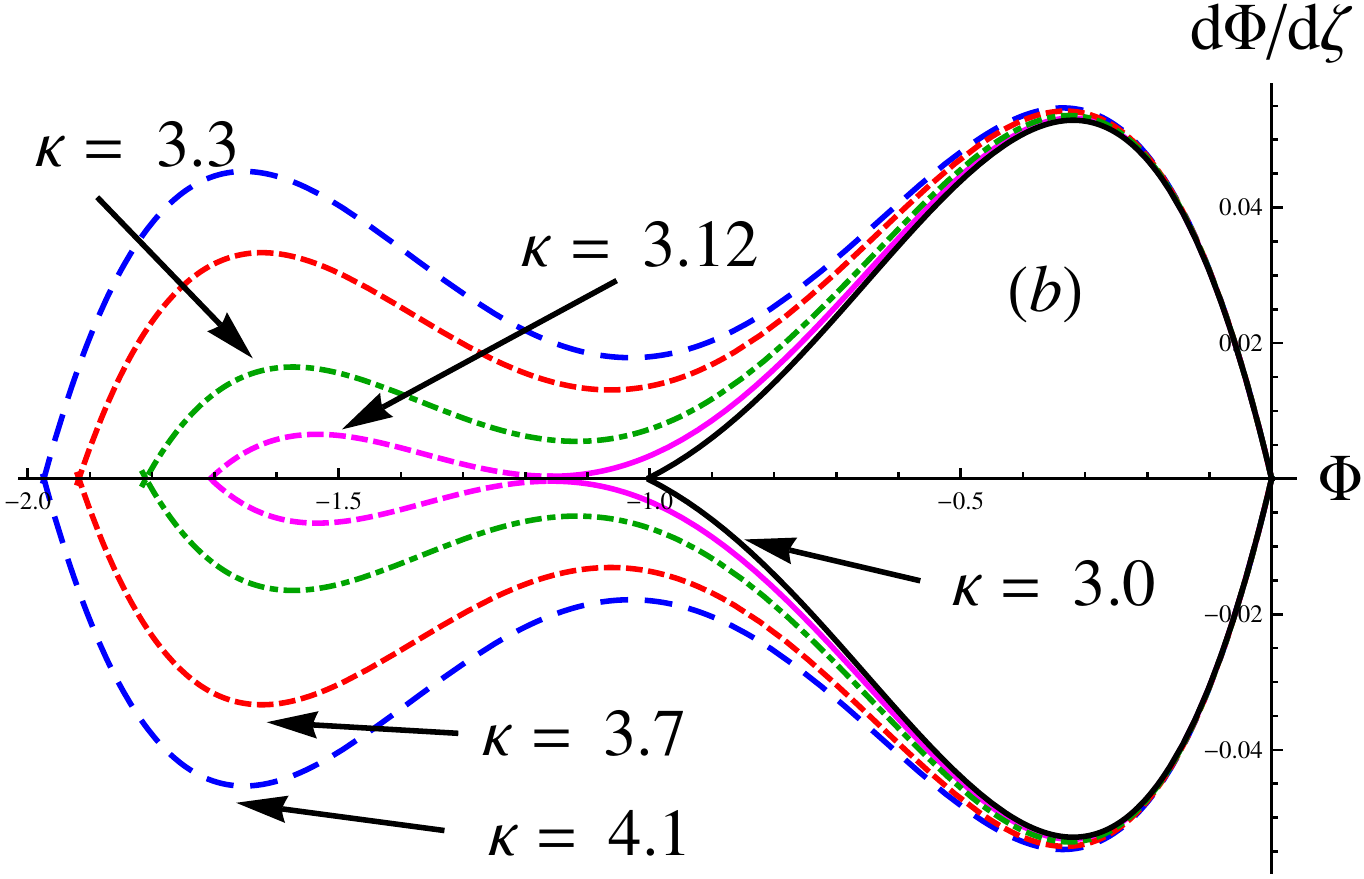} \\
\caption{(a) : Plot of $V(\Phi)$ vs $\Phi$, for different values of $\kappa$.  ~~ (b) : Plot of $d \Phi / d \zeta$ vs $\Phi$ for corresponding values of $\kappa$. ~~ $\kappa = 4.1$ for dashed blue, $\kappa = 3.7$ for dotted red, $\kappa = 3.3$ for dotdashed green, $\kappa = 3.12$ for magenta, and $\kappa = 3.0 $ for solid black curves. Other plasma parameters are $\mu _e = 0.7, \ \mu _d = 0.19, \  \alpha = 4, \ v = 0.39$ }
\ec
\efg

\bfg
\bc
\includegraphics[width = 7 cm]{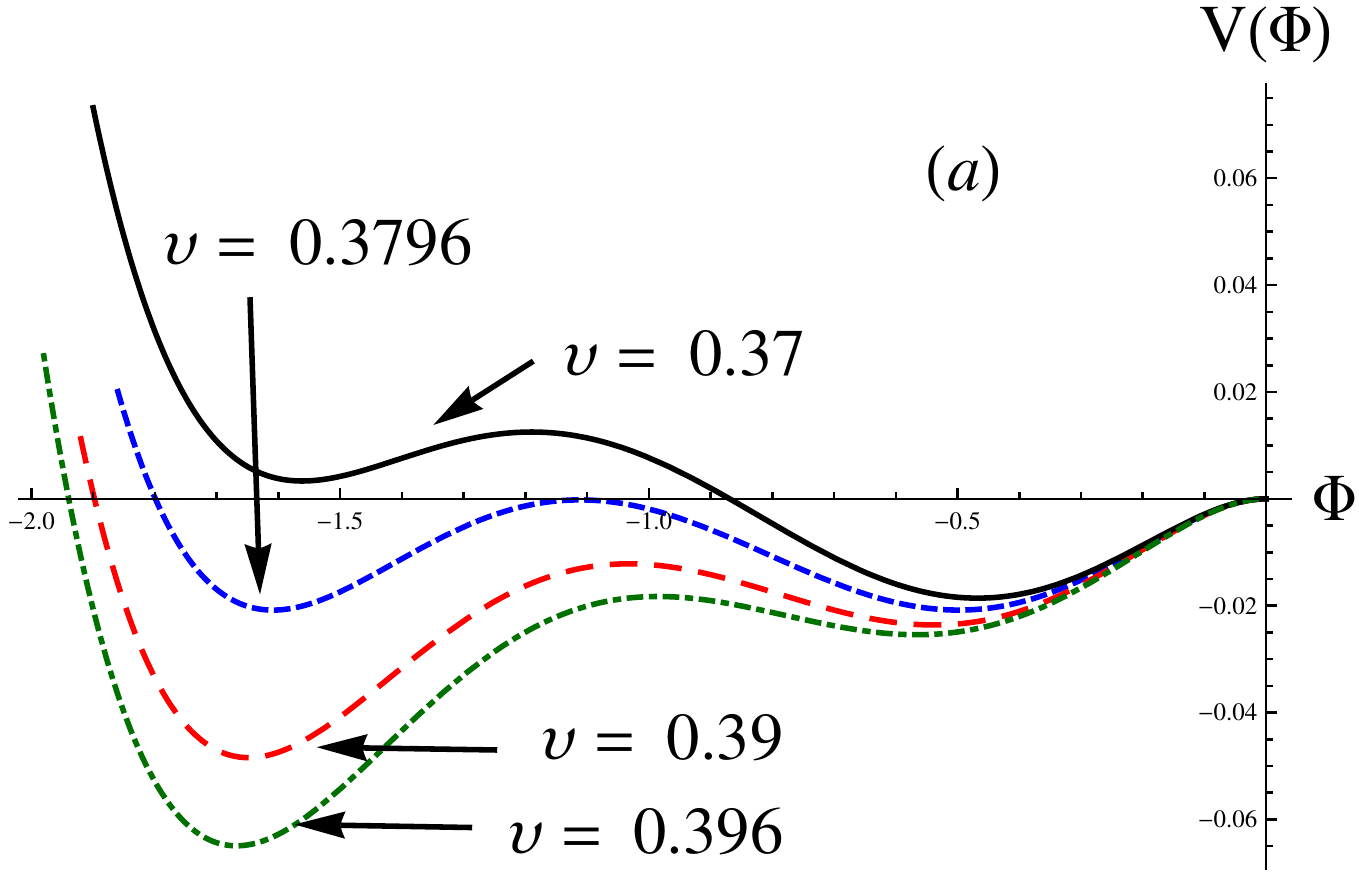} ~~ \includegraphics[width = 7 cm]{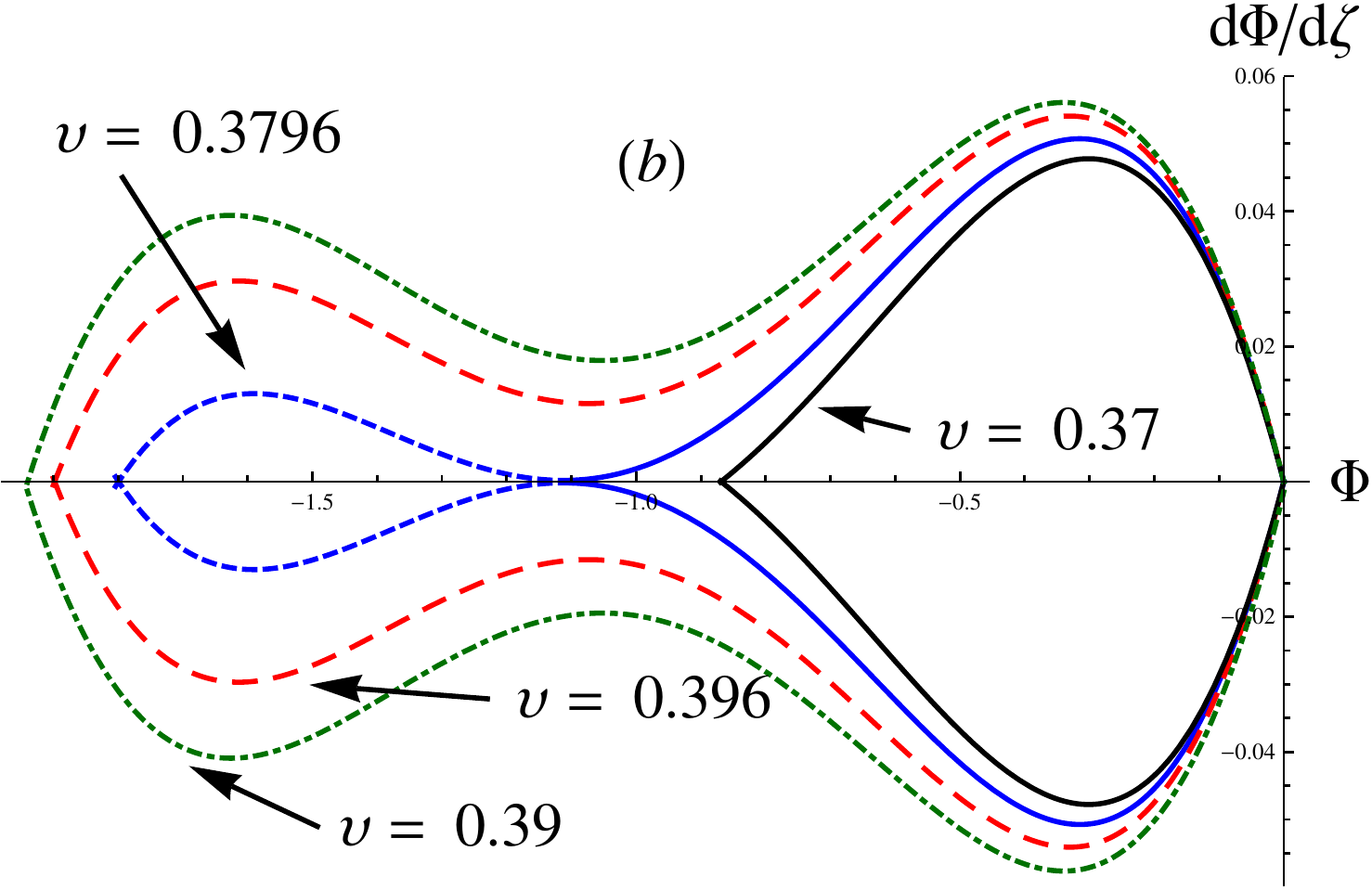} \\
\caption{(a) : Plot of $V(\Phi)$ vs $\Phi$, for different values of $v$.  ~~ (b) : Plot of $d \Phi / d \zeta$ vs $\Phi$ for corresponding different values of $v$. The black curves for $v=0.37$ depict solilton, while the dashed red and dotdashed green curves for $v=0.39$ and $0.396$, respectively, show supersolitons. The blue dotted curves at $v=0.3796$ show double layer. Other plasma parameters are $\mu _e = 0.7, \ \mu _d = 0.19, \  \alpha = 4, \kappa = 3.6$ }
\ec
\efg

\section{Conclusions and Remarks :}

To conclude, in this study we have investigated the propagation of small amplitude nonlinear Dust Ion Acoustic waves in a four-component plasma, consisting of Maxwellian negative ions, cold mobile positive ions, $\kappa$-distributed electrons and positively charged dust grains, as those found in astrophysical environments and also in  laboratory experiments involving ultrasmall devices. Such studies have attracted attention in recent times after the theoretical prediction of supersolitary waves in 2012, and subsequent data received from space plasmas. The presence of charged dust grains in plasma not only modifies the existing plasma wave spectra but also introduces a number of novel eigenmodes, such as the Dust-Ion-Acoustic (DIAWs) waves. These DIAWs play an important role in understanding different types of collective processes in such plasmas.

 Based on the Sagdeev pseudopotential approach, we have studied small amplitude nonlinear structures, in the framework of reductive pertubation theory, with an aim to see if the system supports supersolitons in some parameter domain. After deriving the pseudopotential, we have plotted the same using Mathematica, for various plasma compositions. We have also plotted the corresponding phase portraits, to have a better insight into the nonlinear wave propagation. Within a narrow range of plasma parameters, we observe three types of nonlinear waves --- solitons, double layers and supersolitons. Our plots for the wave solutions as well as electric field profiles agree with our plots for the pseudopotential and phase portraits. Effectively, our present study reveals the following features : 

\begin{itemize}
	\item The four-component plasma studied in this work supports three types of Dust Ion Acoustic waves --- solitons, double layers and supersolitons --- within a very narrow domain of plasma composition, such that the parameters strictly obey certain constraints.
	
	\item Linearization technique fails to reveal the rich structure of the nonlinear waves. The linearized waves obey a very simple dispersion relation, as shown in Fig. 1. For very small wave vector, the frequency, $\omega$, of the linear waves varies linearly with $k$, and finally reaches unity. The spectral index $\kappa$ has very little influence on the linear dispersion relation --- for very low values of $\kappa$, the wave frequency $\omega$ reaches unity at a slightly greater value of $k$.

	\item From each figure depicting the Sagdeev pseudopotential $V(\Phi)$ --- viz., Figures 2(a), 5(a) and 6(a) --- it is observed that $V(\Phi)$ always passes through the origin, which is the unstable equilibrium
	point.
	
	\item As the dust grain concentration decreases, the soliton solutions transform to double layer at some critical value. Beyond that, the solutions are supersolitons, whose amplitudes increase as the dust concentration decreases further, but within permissible limits. These are shown graphically in Fig. 2.
	
	\item In contrast, for low Mach number values, the solutions are standard solitons. By  increasing the Mach number, keeping all other plasma parameters fixed, a double layer appears at some critical value, say $v_c$. On further increasing the Mach number, the nonlinear structures appear as supersolitons, whose amplitudes increase with increasing Mach number. Fig. 5 validates this observation.
	
	\item Similar is the case with $\kappa$ values --- Fig. 6. For small spectral index $\kappa$, the solutions are solitons. Increasing the value of $\kappa$ keeping other plasma parameters unchanged, a double layer appears at a critical value, say $\kappa _c$. Increasing the spectral index further produces supersolitons, of increased amplitude.
	
	\item The supersoliton structure is easily identified from the phase portraits plotted in Figures 2(b), 5(b) and 6(b). Their trajectories are closed ones, enveloping the separatrix (trajectory of the double layer). This indicates that their toal energy is above a certain potential height, and their amplitude cannot be smaller than that of the separatrix.
	
	\item Fig. 3 clearly distinguishes a soliton from a supersoliton in terms of width, amplitude and shape, while the extra wiggle signature in the electric field of supersolitons is shown in Fig. 4, thus proving beyond doubt that in the system studied here DIAWs do propagate as supersolitons in some particular combination of plasma parameters.
	
\end{itemize}

Thus within a very small parameter domain, such that the conditions for the existence of supersolitons are satisfied, the interplay between the different plasma parameters play a very crucial role in determining the nature of the Dust Ion Acoustic waves in the four-component plasma studied here. In particular, depending on the plasma parameter values, three different types of nonlinear waves are observed --- viz., solitons, double layers and supersolitons. However, the rich structure is not reflected in the linearized waves.


\vspace{.3cm}

The present study may be useful in the study of astrophysical and space plasmas where dust grains play a significant role, and also in laboratory experiments involving microelectronic devices. As an extension to this work, we plan to incorporate other factors like viscosity, collisions, etc. in the near future.

\section{Acknowledgment}

 One of the authors (A.S.) thanks the Department of Science and Technology, Govt. of India, for financial assistance, through its grant SR/WOS-A/PM-14/2016.

\end{document}